\documentclass{article}
\usepackage{ijcai19}

\usepackage{times}
\usepackage{soul}
\usepackage{url}
\usepackage[hidelinks]{hyperref}
\usepackage[utf8]{inputenc}
\usepackage[small]{caption}
\usepackage{graphicx}
\usepackage{amsmath}
\usepackage{amsthm}
\usepackage{amsfonts}
\usepackage{amssymb}
\usepackage{booktabs}
\usepackage{algorithm}
\usepackage{algorithmic}
\usepackage[font=footnotesize]{subcaption}
\usepackage[normalem]{ulem}
\usepackage{multicol}
\usepackage{dblfloatfix}
\newtheorem{theorem}{Theorem}[section]

\newtheorem{definition}[theorem]{Definition}

\title{Evidence Propagation and Consensus Formation in Noisy Environments}

\author{
Michael Crosscombe$^1$\footnote{Contact author.}\and
Jonathan Lawry$^1$\and
Palina Bartashevich$^2$\\
\affiliations
$^1$University of Bristol, United Kingdom\\
$^2$Otto von Guericke University Magdeburg, Germany\\
\emails
\{m.crosscombe, j.lawry\}@bristol.ac.uk\\
palina.bartashevich@ovgu.de}

\begin{document}

\maketitle

\blfootnote{This work is published in the proceedings of the 13th International Conference on Scalable Uncertainty Management (SUM 2019), Compi\`egne, France, December 16--18, 2019.}

\begin{abstract}
    We study the effectiveness of consensus formation in multi-agent systems where there is both belief updating based on direct evidence and also belief combination between agents.
    In particular, we consider the scenario in which a population of agents collaborate on the best-of-$n$ problem where the aim is to reach a consensus about which is the best (alternatively, \emph{true}) state from amongst a set of states, each with a different quality value (or level of evidence). Agents' beliefs are represented within Dempster-Shafer theory by mass functions and we investigate the macro-level properties of four well-known belief combination operators for this multi-agent consensus formation problem: Dempster's rule, Yager's rule, Dubois \& Prade's operator and the averaging operator. The convergence properties of the operators are considered and simulation experiments are conducted for different evidence rates and noise levels.
    Results show that a combination of updating on direct evidence and belief combination between agents results in better consensus to the best state than does evidence updating alone. We also find that in this framework the operators are robust to noise.
    Broadly, Yager's rule is shown to be the better operator under various parameter values, i.e. convergence to the best state, robustness to noise, and scalability.
\end{abstract}


\section{Introduction and background}
\label{section:introduction}

Agents operating in noisy and complex environments receive evidence from a variety of different sources, many of which will be at least partially inconsistent. In this paper we investigate the interaction between two broad categories of evidence: ({\romannumeral 1}) direct evidence from the environment, and ({\romannumeral 2}) evidence received from other agents with whom an agent is interacting or collaborating to perform a task. For example, robots engaged in a search and rescue mission will receive data directly from sensors as well as information from other robots in the team.
Alternatively, software agents can have access to online data as well as sharing data with other agents.

The efficacy of combining these two types of evidence in multi-agent systems has been studied from a number of different perspectives. In social epistemology~\cite{Douven2011} has argued that agent-to-agent communications has an important role to play in propagating locally held information widely across a population. For example, interaction between scientists facilitates the sharing of experimental evidence. Simulation results are then presented which show that a combination of direct evidence and agent interaction, within the Hegselmann-Krause opinion dynamics model~\cite{hegselmann02}, results in faster convergence to the true state than updating based solely on direct evidence. A probabilistic model combining Bayesian updating and probability pooling of beliefs in an agent-based system has been proposed in~\cite{Lee2018}. In this context it is shown that combining updating and pooling leads to faster convergence and better consensus than Bayesian updating alone. An alternative methodology exploits three-valued logic to combine both types of evidence~\cite{Crosscombe2016} and has been effectively applied to distributed decision-making in swarm robotics~\cite{Crosscombe2017}.

In this current study we exploit the capacity of Dempster-Shafer theory (DST) to fuse conflicting evidence in order to investigate how direct evidence can be combined with a process of iterative belief aggregation in the context of the best-of-$n$ problem. The latter refers to a general class of problems in distributed decision-making~\cite{Parker2009,Valentini2017} in which a population of agents must collectively identify which of $n$ alternatives is the correct, or best, choice. These alternatives could correspond to physical locations as, for example, in a search and rescue scenario, different possible states of the world, or different decision-making or control strategies. Agents receive direct but limited feedback in the form of quality values associated with each choice, which then influence their beliefs when combined with those of other agents with whom they interact.
It is not our intention to develop new operators in DST nor to study the axiomatic properties of particular operators at the local level (see~\cite{Dubois2016} for an overview of such properties). Instead, our motivation is to study the macro-level convergence properties of several established operators when applied \emph{iteratively} by a population of agents, over long timescales, and in conjunction with a process of evidential updating, i.e., updating beliefs based on evidence.

An outline of the remainder of the paper is as follows. In Section~\ref{section:model} we give a brief introduction to the relevant concepts from DST and summarise its previous application to dynamic belief revision in agent-based systems. Section~\ref{section:bestn} introduces a version of the best-of-$n$ problem exploiting DST measures and combination operators. In Section~\ref{section:fixedpoint} we then give the fixed point analysis of a dynamical system employing DST operators so as to provide insight into the convergence properties of such systems. In Section~\ref{section:simulation-experiments} we present the results from a number of agent-based simulation experiments carried out to investigate consensus formation in the best-of-$n$ problem under varying rates of evidence and levels of noise. Finally, Section~\ref{section:conclusion} concludes with some discussion.

\section{An Overview of Dempster-Shafer Theory}
\label{section:model}


In this section we introduce relevant concepts from Dempster-Shafer theory (DST)~\cite{Dempster1967,Shafer1976}, including four well-known belief combination operators.

\begin{definition}{ Mass function (or agent's belief)}

Given a set of states or frame of discernment $\mathbb{S} = \{s_1, ..., s_n\}$, let $2^{\mathbb{S}}$ denote the power set of $\mathbb{S}$. An agent's belief is then defined by a basic probability assignment or mass function $m : 2^{\mathbb{S}} \rightarrow [0,1]$, where $m(\emptyset) = 0$ and $\sum_{A \subseteq \mathbb{S}} m(A) = 1$. The mass function then characterises a belief and a plausibility measure defined on $2^{\mathbb{S}}$ such that for $A \subseteq \mathbb{S}:$
\begin{gather*}
    Bel(A)=\sum_{B\subseteq A} m(B) \text{ and } Pl(A)=\sum_{B:B\cap A \neq \emptyset} m(B)
\end{gather*}
and hence where $Pl(A)=1-Bel(A^c)$.
\end{definition}

A number of operators have been proposed in DST for combining or fusing mass functions~\cite{Smets2007}. In this paper we will compare in a dynamic multi-agent setting the following operators: Dempster's rule of combination (\textbf{DR})~\cite{Shafer1976}, Dubois \& Prade's operator (\textbf{D\&P})~\cite{Dubois1988}, Yager's rule (\textbf{YR})~\cite{Yager1992}, and a simple averaging operator (\textbf{AVG}). The first three operators all make the assumption of independence between the sources of the evidence to be combined but then employ different techniques for dealing with the resulting inconsistency. DR uniformly reallocates the mass associated with non-intersecting pairs of sets to the overlapping pairs, D\&P does not re-normalise in such cases but instead takes the union of the two sets, while YR reallocates all inconsistent mass values to the universal set $\mathbb{S}$. These four operators were chosen based on several factors: the operators are well established and have been well studied, they require no additional information about individual agents, and they are computationally efficient at scale (within the limits of DST). 

\begin{definition}{Combination operators}
\label{dfn:operators}

Let $m_1$ and $m_2$ be mass functions on $2^{\mathbb{S}}$. Then the combined mass function $m_1 \odot m_2$ is a function $m_1 \odot m_2 \colon 2^{\mathbb{S}} \rightarrow [0,1]$ such that for $\emptyset \neq A, B, C \subseteq \mathbb{S}:$
\begin{equation*}
    \begin{split}
        \textbf{\emph{(DR)} } m_1 \odot m_2 (C)&= \frac{1}{1 - K} \sum_{A \cap B = C \neq \emptyset} m_1(A) \cdot m_2(B), \\
        \textbf{\emph{(D\&P)} } m_1 \odot m_2 (C) &= \sum_{A \cap B = C \neq \emptyset} m_1(A) \cdot m_2(B)\\
        &+\sum_{\substack{A \cap B = \emptyset, \\ A \cup B = C}} m_1(A) \cdot m_2(B), \\
        \textbf{\emph{(YR)} } m_1 \odot m_2 (C)&= \sum_{A \cap B = C \neq \emptyset} m_1(A) \cdot m_2(B) \mbox{ if } C \neq \mathbb{S},\\
        m_1 \odot m_2(\mathbb{S})&= m_1(\mathbb{S}) \cdot m_2(\mathbb{S})+K, \\
        \textbf{\emph{(AVG)} } m_1 \odot m_2 (C) &= \frac{1}{2} \left( m_1(C) + m_2(C)\right),
    \end{split}
\end{equation*}
where $K$ is associated with conflict, i.e., $K=\sum_{A \cap B = \emptyset} m_1(A) \cdot m_2(B)$.
\end{definition}



In the agent-based model of the best-of-$n$ problem, proposed in Section~\ref{section:bestn}, agents are required to make a choice as to which of $n$ possible states they should investigate at any particular time. To this end we utilise the notion of \emph{pignistic distribution} proposed by Smets and Kennes~\cite{Smets1994}. 

\begin{definition}{Pignistic distribution}

    For a given mass function $m$, the corresponding pignistic distribution on $\mathbb{S}$ is a probability distribution obtained by reallocating the mass associated with each set $A \subseteq \mathbb{S}$ uniformly to the elements of that set, i.e., $s_i \in A$, as follows:
    \begin{displaymath}
        P(s_i|m)= \sum_{A:s_i \in A} \frac{m(A)}{|A|}.
    \end{displaymath}
	\label{def:pignistic}
\end{definition}

DST has been applied to multi-agent dynamic belief revision in a number of ways. For example, \cite{Dabarera2014} and \cite{Wikramarathne2014} investigate belief revision where agents update their beliefs by taking a weighted combination of conditional belief values of other agents using Fagin-Halpern conditional belief measures. These measures are motivated by the probabilistic interpretation of DST according to which a belief and plausibility measure are characterised by a set of probability distributions on $\mathbb{S}$.
Several studies~\cite{Cho2014,Crosscombe2016,Lu2015} have applied a three-valued version of DST in multi-agent simulations.
This corresponds to the case in which there are two states, $\mathbb{S}=\{s_1,s_2\}$, one of which is associated with the truth value \emph{true} (e.g., $s_1$), one with \emph{false} ($s_2$), and where the set $\{s_1,s_2\}$ is then taken as corresponding to a third truth state representing \emph{uncertain} or \emph{borderline}.
One such approach based on subjective logic \cite{Cho2014} employs the combination operator proposed in \cite{Josang2002}. Another~\cite{Lu2015} uses Dempster's rule applied to combine an agent's beliefs with an aggregate of those of her neighbours. Similarly, \cite{Crosscombe2016} uses Dubois \& Prade's operator for evidence propagation.
Other relevant studies include \cite{Kanjanatarakul2017} in which Dempster's rule is applied across a network of sparsely connected agents.

With the exception of~\cite{Crosscombe2016}, and only for \emph{two} states, none of the above studies considers the interaction between direct evidential updating and belief combination. The main contribution of this paper is therefore to provide a detailed and general study of DST applied to dynamic multi-agent systems in which there is both direct evidence from the environment and belief combination between agents with partially conflicting beliefs.
In particular, we will investigate and compare the consensus formation properties of the four combination operators (Definition~\ref{dfn:operators}) when applied to the best-of-$n$ problem.


\section{The Best-of-\texorpdfstring{$n$}{n} Problem within DST}
\label{section:bestn}
Here we present a formulation of the best-of-$n$ problem within the DST framework. We take the $n$ choices to be the states $\mathbb{S}$. Each state $s_i \in \mathbb{S}$ is assumed to have an associated quality value $q_i \in [0,1]$ with 0 and 1 corresponding to minimal and maximal quality, respectively. Alternatively, we might interpret $q_i$ as quantifying the level of available evidence that $s_i$ corresponds to the true state of the world.

In the best-of-$n$ problem agents explore their environment and interact with each other with the aim of identifying which is the highest quality (or true) state. Agents sample states and receive evidence in the form of the quality $q_i$, so that in the current context evidence $E_i$ regarding state $s_i$ takes the form of the following mass function;
\begin{gather*}
    m_{E_i}=\{s_i\}:q_i, \ \mathbb{S}:1-q_i.
\end{gather*}
Hence, $q_i$ is taken as quantifying both the evidence directly in favour of $s_i$ provided by $E_i$, and also the evidence directly against any other state $s_j$ for $j \neq i$. Given evidence $E_i$ an agent updates its belief by combining its current mass function $m$ with $m_{E_i}$ using a combination operator so as to obtain the new mass function given by $m \odot m_{E_i}$.

A summary of the process by which an agent might obtain direct evidence in this model is then as follows. Based on its current mass function $m$, an agent stochastically selects a state $s_i \in \mathbb{S}$ to investigate\footnote{We utilise roulette wheel selection; a proportionate selection process.}, according to the pignistic probability distribution for $m$ as given in Definition~\ref{def:pignistic}. More specifically, it will update $m$ to $m \odot m_{E_i}$ with probability $P(s_i|m)\times r$ for $i=1,\ldots,n$ and leave its belief unchanged with probability $(1-r)$, where $r \in [0,1]$ is a fixed evidence rate quantifying the probability of finding evidence about the state that it is currently investigating. In addition, we also allow for the possibility of noise in the evidential updating process. This is modelled by a random variable $\epsilon\sim \mathcal{N}(0,\,\sigma^{2})$ associated with each quality value. In other words, in the presence of noise the evidence $E_i$ received by an agent has the form:
\begin{gather*}
    m_{E_i}=\{s_i\}:q_i+\epsilon, \mathbb{S}:1-q_i-\epsilon,
\end{gather*}
where if $q_i + \epsilon <0$ then it is set to $0$, and if $q_i + \epsilon >1$ then it is set to $1$.
Overall, the process of updating from direct evidence is governed by the two parameters, $r$ and $\sigma$, quantifying the availability of evidence and the level of associated noise, respectively.

In addition to receiving direct evidence we also include belief combination between agents in this model. This is conducted in a pairwise symmetric manner in which two agents are selected at random to combine their beliefs, with both agents then adopting this combination as their new belief, i.e., if the two agents have beliefs $m_1$ and $m_2$, respectively, then they both replace these with $m_1 \odot m_2$.
However, in the case that agents are combining their beliefs under Dempster's rule and that their beliefs are completely inconsistent, i.e., when $K=1$ (see Definition~\ref{dfn:operators}), then they do not form consensus and the process moves on to the next iteration.

In summary, during each iteration both processes of evidential updating and consensus formation take place\footnote{Due to the possibility of rounding errors occurring as a result of the multiplication of small numbers close to $0$,
we renormalise the mass function that results from each each process.}.
However, while every agent in the population has the potential to update its own belief, provided that it successfully receives a piece of evidence, the consensus formation is restricted to a \textit{single pair} of agents for each iteration. That is, we assume that only two agents in the whole population are able to communicate and combine their beliefs during each iteration.




\section{Fixed Point Analysis}
\label{section:fixedpoint}
In the following, we provide an insight into the convergence properties of the dynamical system described in Section~\ref{section:model}.
Consider an agent model in which at each time step $t$ two agents are selected at random to combine their beliefs from a population of $k$ agents $\mathcal{A}=\{a_1 \ldots, a_k\}$ with beliefs quantified by mass functions $m_i^t:i=1, \ldots, k$. For any $t$ the state of the system can be represented by a vector of mass functions $\langle m_1^t, \ldots, m_k^t \rangle$. Without loss of generality, we can assume that the updated state is then $\langle m_1^{t+1},m_2^{t+1}, \ldots, m_k^{t+1}\rangle=\langle m_1^t \odot m_2^t, m_1^t \odot m_2^t, m_3^t, \ldots, m_k^t \rangle$.
Hence, we have a dynamical system characterised by the following mapping:
\begin{gather*}
\langle m_1^t, \ldots, m_k^t\rangle \rightarrow \langle m_1^t \odot m_2^t, m_1^t \odot m_2^t, m_3^t, \ldots, m_k^t \rangle.
\end{gather*}
The fixed points of this mapping are those for which $m_1^t=m_1^t \odot m_2^t$ and $m_2^t=m_1^t \odot m_2^t$. This requires that $m_1^t=m_2^t$ and hence the fixed point of the mapping are the fixed points of the operator, i.e., those mass functions $m$ for which $m \odot m=m$.

\begin{figure*}[hb]
\begin{center}
\begin{subfigure}{.7\textwidth}
    \includegraphics[width=1\textwidth]{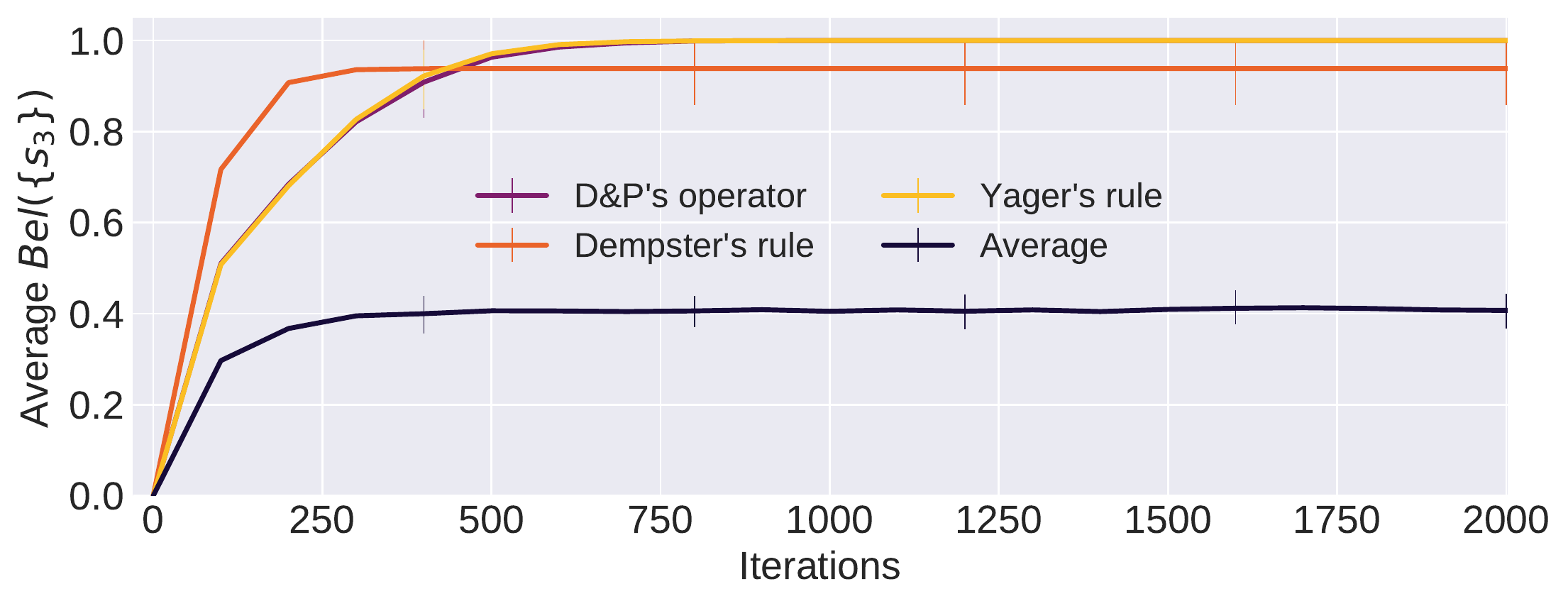}
\end{subfigure}
\caption{Average $Bel(\{s_3\})$ plotted against iteration $t$ with $r=0.05$ and $\sigma=0.1$. Comparison of all four operators with error bars displaying the standard deviation.}
\label{fig:traj}
\end{center}
\end{figure*}

Let us analyse in detail the fixed points for the case in which there are $3$ states $\mathbb{S}=\{s_1,s_2,s_3\}$.
Let $m=\{s_1,s_2,s_3\}:x_7, \{s_1,s_2\}:x_4, \{s_1,s_3\}:x_5,\{s_2,s_3\}:x_6, \{s_1\}:x_1,  \{s_2\}:x_2, \{s_3\}:x_3$ represent a general mass function defined on this state space and where without loss of generality we take $x_7=1-x_1-x_2-x_3-x_4-x_5-x_6$. For Dubois \& Prade's operator the constraint that $m \odot m=m$ generates the following simultaneous equations.
\begin{gather*}
    x_1^2 + 2x_1 x_4 + 2 x_1 x_5 + 2 x_1 x_7 + 2x_4 x_5 = x_1\\
    x_2^2 + 2 x_2 x_4 + 2x_2x_6 + 2 x_2 x_7 + 2x_4 x_6 = x_2\\
    x_3^2 + 2 x_3 x_5 + 2 x_3 x_6 + 2 x_3 x_7 + 2 x_5 x_6 = x_3 \\
    x_4^2 + 2 x_1 x_2 + 2 x_4 x_7=x_4\\
    x_5^2 + 2 x_1 x_3 + 2x_5 x_7=x_5\\
    x_6^2+ 2 x_2x_3 + 2 x_6 x_7=x_6
\end{gather*}
The Jacobian for this set of equations is given by:
\begin{gather*}
    \mathbf{J}=\left( \frac{\partial}{\partial x_j} m \odot m(A_i)  \right),
\end{gather*}
where $A_1=\{s_1\}, A_2=\{s_2\}, A_3=\{s_3\}, A_4=\{s_1,s_2\}, \ldots$
The stable fixed points are those solutions to the above equations for which the eigenvalues of the Jacobian evaluated at the fixed point lie within the unit circle on the complex plane. In this case the only stable fixed points are the mass functions $\{s_1\}:1$, $\{s_2\}:1$ and $\{s_3\}:1$. In other words, the only stable fixed points are those for which agents' beliefs are both certain and precise. That is where for some state $s_i \in \mathbb{S}$, $Bel(\{s_i\})=Pl(\{s_i\})=1$. The stable fixed points for Dempster's rule and Yager's rule are also of this form. The averaging operator is idempotent and all mass functions are unstable fixed points.

The above analysis concerns agent-based systems applying a combination in order to reach consensus. However, we have yet to incorporate evidential updating into this model. As outlined in Section~\ref{section:bestn}, it is proposed that each agent investigates a particular state $s_i$ chosen according to its current beliefs using the pignistic distribution. With probability $r$ this will result in an update to its beliefs from $m$ to $m \odot m_{E_i}$. Hence, for convergence it is also required that agents only choose to investigate states for which $m \odot  m_{E_i}=m$. Assuming $q_i>0$, then there is only one such fixed point corresponding to $m=\{s_i\}:1$. Hence, the consensus driven by belief combination as characterised by the above fixed point analysis will result in convergence of individual agent beliefs if we also incorporate evidential updating. That is, an agent with beliefs close to a fixed point of the operator, i.e., $m=\{s_i\}:1$, will choose to investigate state $s_i$ with very high probability and will therefore tend to be close to a fixed point of the evidential updating process.

\begin{figure*}[t]
\begin{subfigure}[t]{.48\textwidth}
    \includegraphics[width=1\textwidth]{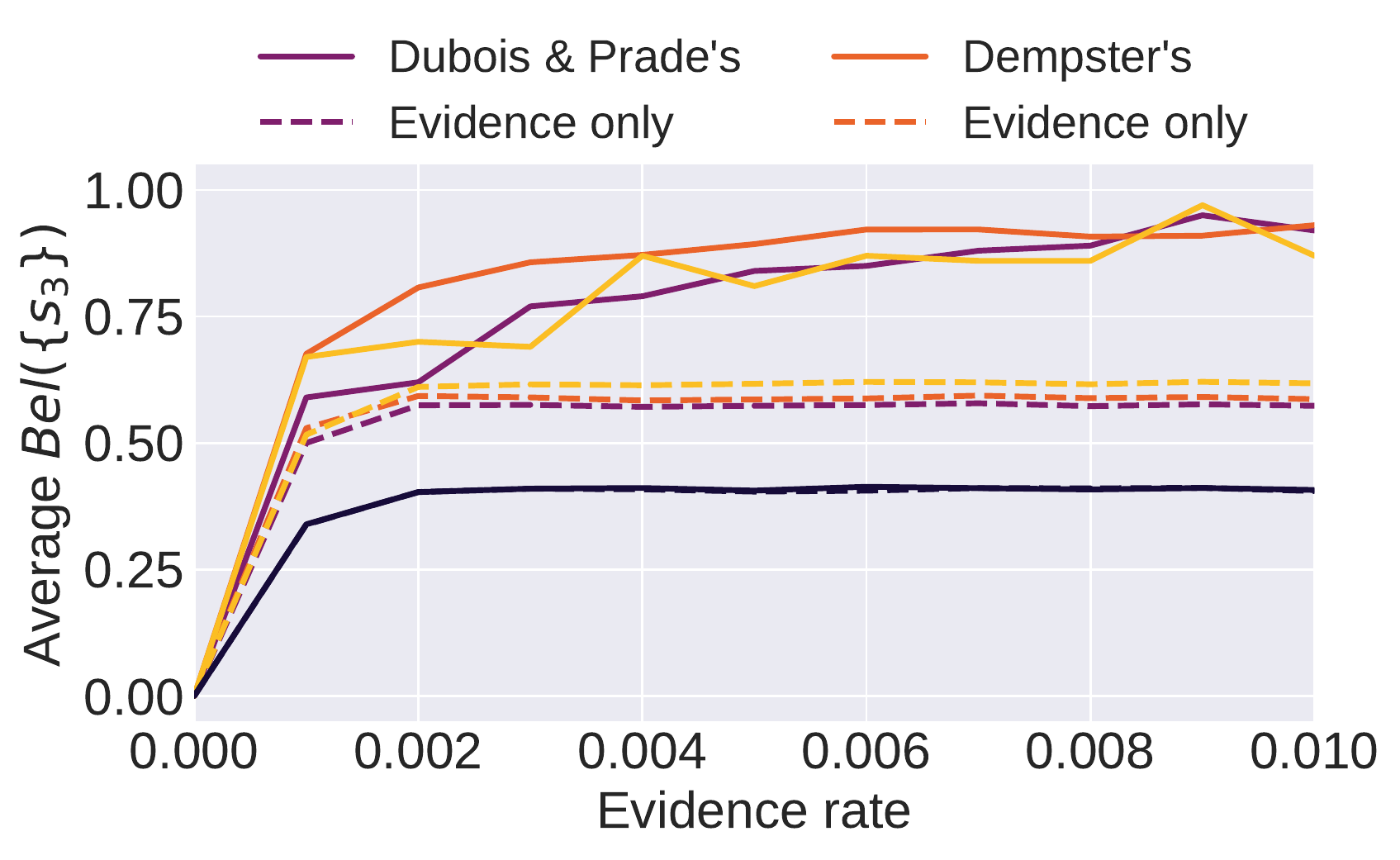}
    \subcaption{Low evidence rates $r \in [0, 0.01]$.}
    \label{fig:evidence_rate_zoomed}
\end{subfigure}\hfill
\begin{subfigure}[t]{.48\textwidth}
    \includegraphics[width=1\textwidth]{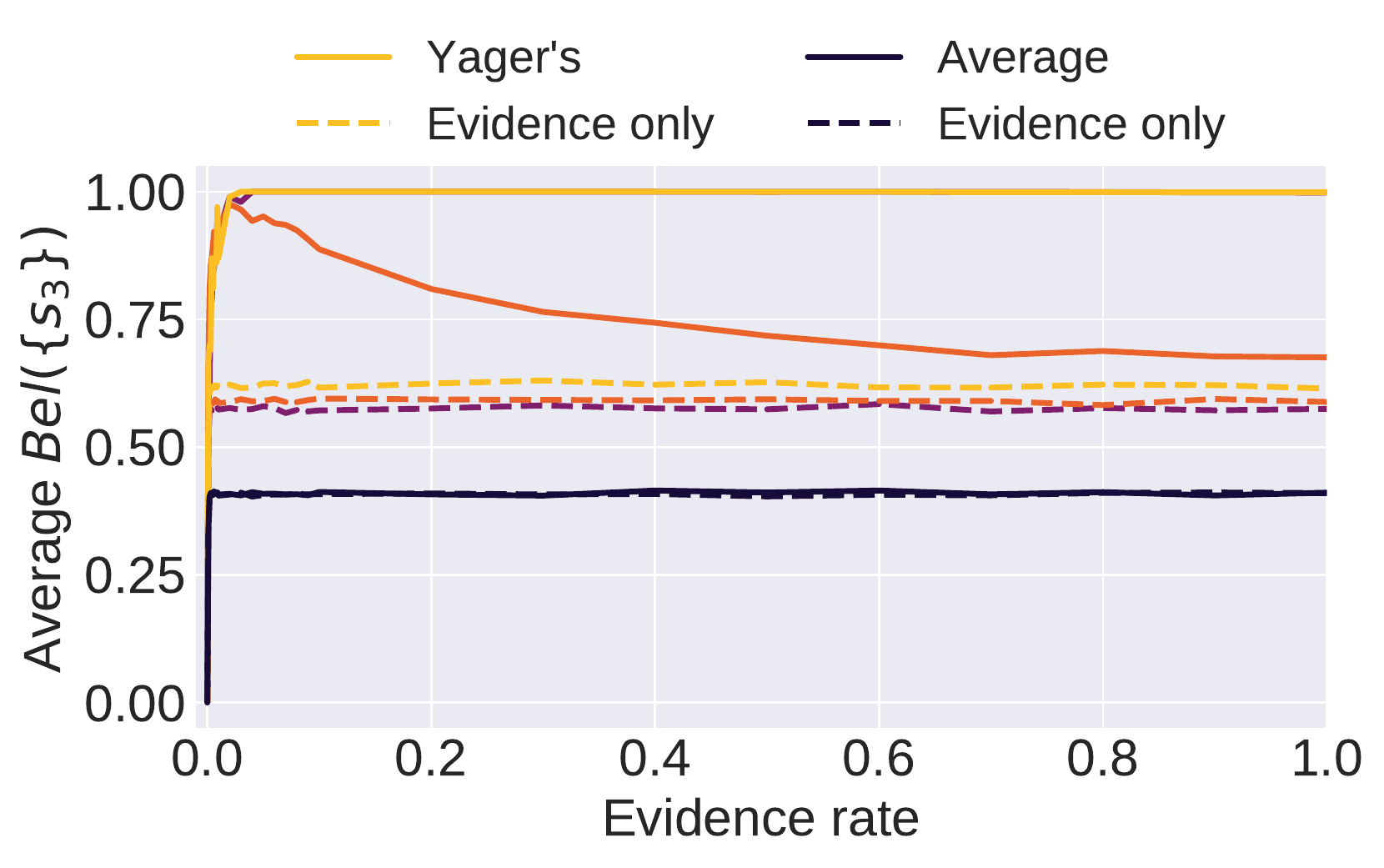}
    \subcaption{All evidence rates $r \in [0, 1]$.}
    \label{fig:evidence_rate_full}
\end{subfigure}\hfill
\caption{Average $Bel(\{s_3\})$ for evidence rates $r \in [0, 1]$. Comparison of all four operators both with and without belief combination between agents.}
\label{fig:evidence_rate}
\end{figure*}

\begin{figure*}[b]
\begin{center}
\begin{subfigure}{.7\textwidth}
    \includegraphics[width=1\textwidth]{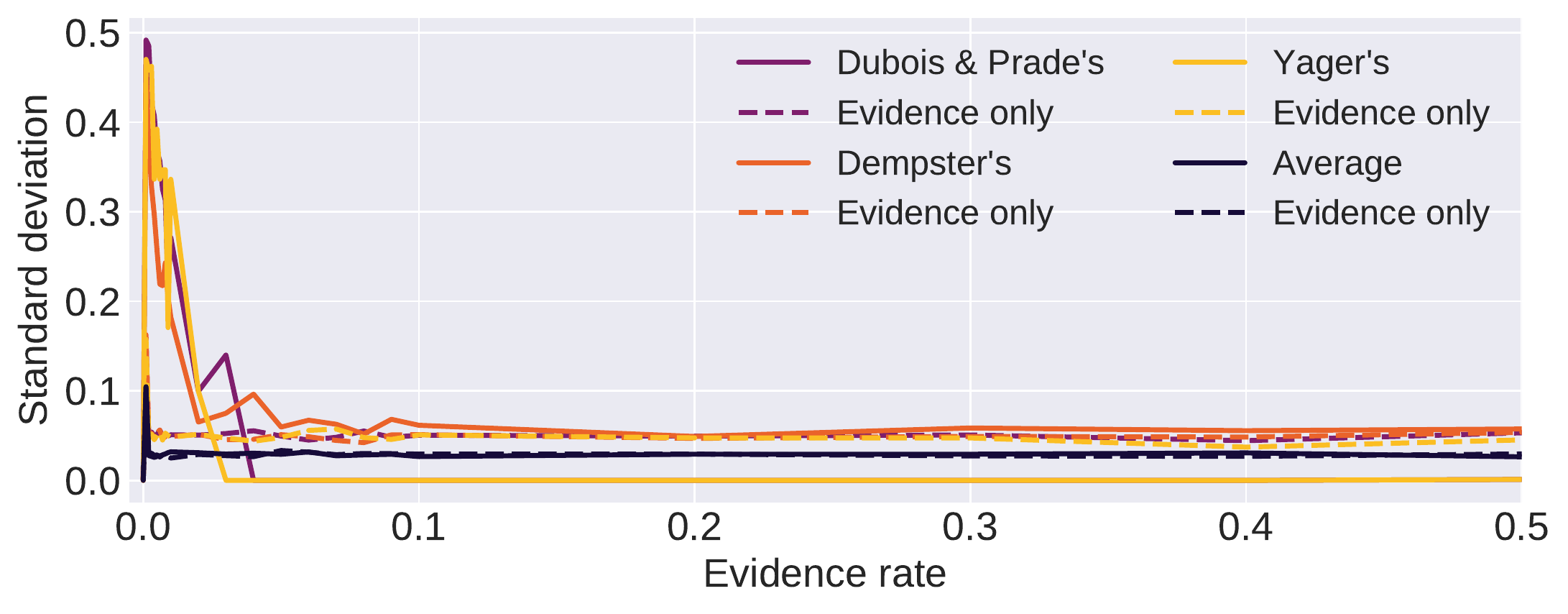}
\end{subfigure}\hfill
\caption{Standard deviation for different evidence rates $r \in [0, 0.5]$. Comparison of all four operators both with and without belief combination between agents.}
\label{fig:std_dev}
\end{center}
\end{figure*}

\section{Simulation Experiments}
\label{section:simulation-experiments}
%

In this section we describe experiments conducted to understand the behaviour of the four belief combination operators in the context of the dynamic multi-agent best-of-$n$ problem introduced in Section~\ref{section:bestn}. We compare their performance under different evidence rates $r$, noise levels $\sigma$, and their scalability for different numbers of states $n$.

\subsection{Parameter Settings}
\label{subsection:parameters}
Unless otherwise stated, all experiments share the following parameter values. We consider a population $\mathcal{A}$ of $k=100$ agents with beliefs initialised so that:
\begin{gather*}
    m_i^0=\mathbb{S}:1 \text{ for } i=1, \ldots, 100.
\end{gather*}
In other words, at the beginning of each simulation every agent is in a state of complete ignorance as represented in DST by allocating all mass to the set of all states $\mathbb{S}$. Each experiment is run for a maximum of $5\,000$ iterations, or until the population converges. Here, convergence requires that the beliefs of the population have not changed for $100$ interactions, where an interaction may be the updating of beliefs based on evidence or the combination of beliefs between agents. For a given set of parameter values the simulation is run $100$ times and results are then averaged across these runs.

Quality values are defined so that $q_i=\frac{i}{n+1}$ for $i=1, \ldots, n$ and consequently $s_n$ is the best state. In the following, $Bel(\{s_n\})$ provides a measure of convergence performance for the considered operators.



\subsection{Convergence Results} 
\label{subsection:results}
Initially we consider the best-of-$n$ problem where $n=3$ with quality values $q_1=0.25$, $q_2=0.5$ and $q_3=0.75$. Figure~\ref{fig:traj} shows belief values for the best state $s_3$ averaged across agents and simulation runs for the evidence rate $r=0.05$ and noise standard deviation $\sigma=0.1$. For both Dubois \& Prade's operator and Yager's rule there is complete convergence to $Bel(\{s_3\})=1$ while for Dempster's rule the average value of $Bel(\{s_3\})$ at steady state is approximately $0.9$.
The averaging operator does not converge to a steady state and instead maintains an average value of $Bel(\{s_3\})$ oscillating around $0.4$.
For all but the averaging operator, at steady state the average belief and plausibility values are equal. This is consistent with the fixed point analysis given for Dubois \& Prade's operator in Section~\ref{section:fixedpoint}, showing that all agents converge to mass functions of the form $m=\{s_i\}:1$ for some state $s_i \in \mathbb{S}$.
Indeed, for both Dubois \& Prade's operator and Yager's rule all agents converge to $m=\{s_3\}:1$, while for Dempster's rule this happens in the large majority of cases. In other words, the combination of updating from direct evidence and belief combination results in agents reaching the certain and precise belief that $s_3$ is the true state of the world.

\begin{figure*}[b]
\centering
\begin{subfigure}[t]{1\textwidth}
    \includegraphics[width=1\textwidth]{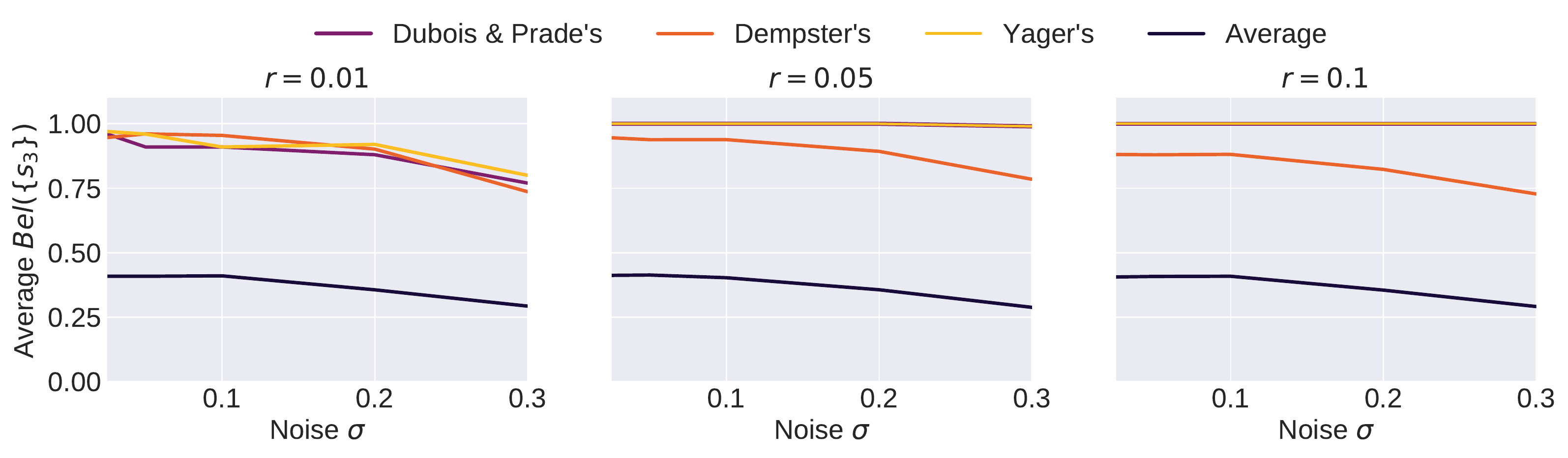}
    \label{fig:noise_er}
\end{subfigure}
\caption{Average $Bel(\{s_3\})$ for all four operators plotted against $\sigma \in [0, 0.3]$ for different evidence rates $r$. Left: $r = 0.01$. Centre: $r = 0.05$. Right: $r = 0.1$.
}
\label{fig:noise}
\end{figure*}

\subsection{Varying Evidence Rates}
\label{ssec:evidence}
In this section we investigate how the rate at which agents receive information from their environment affects their ability to reach a consensus about the true state of the world.

Figures~\ref{fig:evidence_rate_zoomed} and \ref{fig:evidence_rate_full} show average steady state values of $Bel(\{s_3\})$ for evidence rates in the lower range $r \in [0,0.01]$ and across the whole range $r \in [0, 1]$, respectively. For each operator we compare the combination of evidential updating and belief combination (solid lines) with that of evidential updating alone (dashed lines). From Figure~\ref{fig:evidence_rate_zoomed} we see that for low values of $r\leq 0.02$ Dempster's rule converges to higher average values of $Bel(\{s_3\})$ than do the other operators. Indeed, for $0.001 \leq r \leq 0.006$ the average value of $Bel(\{s_3\})$ obtained using Dempster's rule is approximately $10\%$ higher than is obtained using Dubois \& Prade's operator and Yager's rule, and is significantly higher still than that of the averaging operator. However, the performance of Dempster's rule declines significantly for higher evidence rates and for $r >0.3$ it converges to average values for $Bel(\{s_3\})$ of less than $0.8$. At $r=1$, when every agent is receiving evidence at each time step, there is failure to reach consensus when applying Dempster's rule.
Indeed, there is polarisation with the population splitting into separate groups, each certain that a different state is the best.
In contrast, both Dubois \& Prade's operator and Yager's rule perform well for higher evidence rates and for all $r >0.02$ there is convergence to an average value of $Bel(\{s_3\})=1$.
Meanwhile the averaging operator appears to perform differently for increasing evidence rates and instead maintains similar levels of performance for $r > 0.1$.
For all subsequent figures showing steady state results, we do not include error bars as this impacts negatively on readability.
Instead, we show the standard deviation plotted separately against the evidence rate in Figure~\ref{fig:std_dev}. As expected, standard deviation is high for low evidence rates in which the sparsity of evidence results in different runs of the simulation converging to different states. This then declines rapidly with increasing evidence rates.

The dashed lines in Figures~\ref{fig:evidence_rate_zoomed} and \ref{fig:evidence_rate_full} show the values of $Bel(\{s_3\})$ obtained at steady state when there is only updating based on direct evidence. In most cases the performance is broadly no better than, and indeed often worse than, the results which combine evidential updating with belief combination between agents. For low evidence rates where $r<0.1$ the population does not tend to fully converge to a steady state since there is insufficient evidence available to allow convergence. For higher evidence rates under Dempster's rule, Dubois \& Prade's operator and Yager's rule, the population eventually converges on a single state with complete certainty. However, since the average value of $Bel(\{s_3\})$ in both cases is approximately $0.6$ for $r > 0.002$ then clearly convergence is often not to the best state. The averaging operator is not affected by the combined updating method and performs the same under evidential updating alone as it does in conjunction with consensus formation.

Overall, it is clear then that in this formulation of the best-of-$n$ problem combining both updating from direct evidence and belief combination results in much better performance than obtained by using evidential updating alone for all considered operators except the averaging operator. 

\subsection{Noisy Evidence}
\label{ssec:noise}

\begin{figure}[t]
\begin{subfigure}[t]{.48\textwidth}
    \includegraphics[width=1\textwidth]{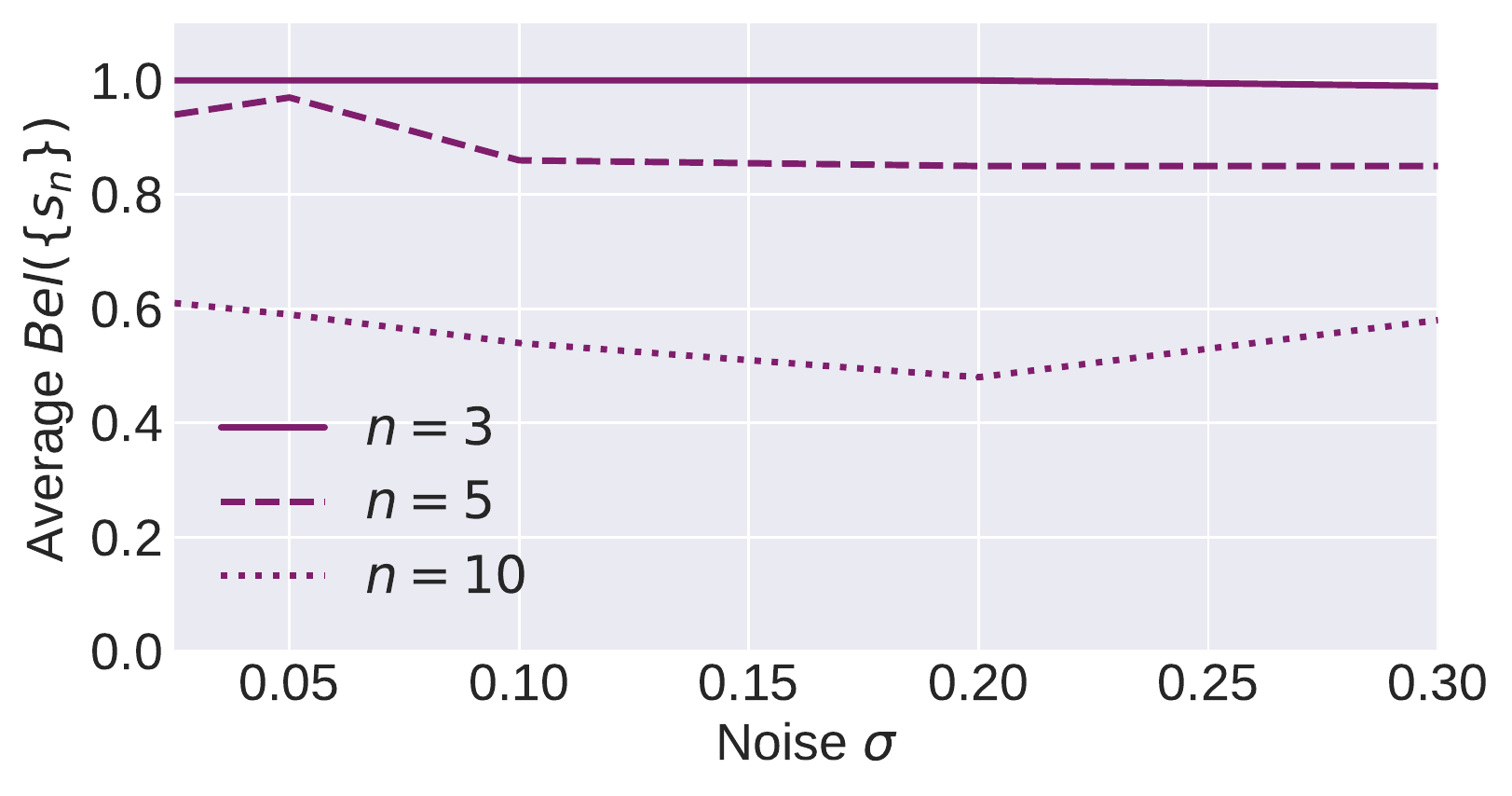}
    \subcaption{Dubois \& Prade's operator.}
    \label{fig:best-of-n-consensus}
\end{subfigure}\hfill
\begin{subfigure}[t]{.48\textwidth}
    \includegraphics[width=1\textwidth]{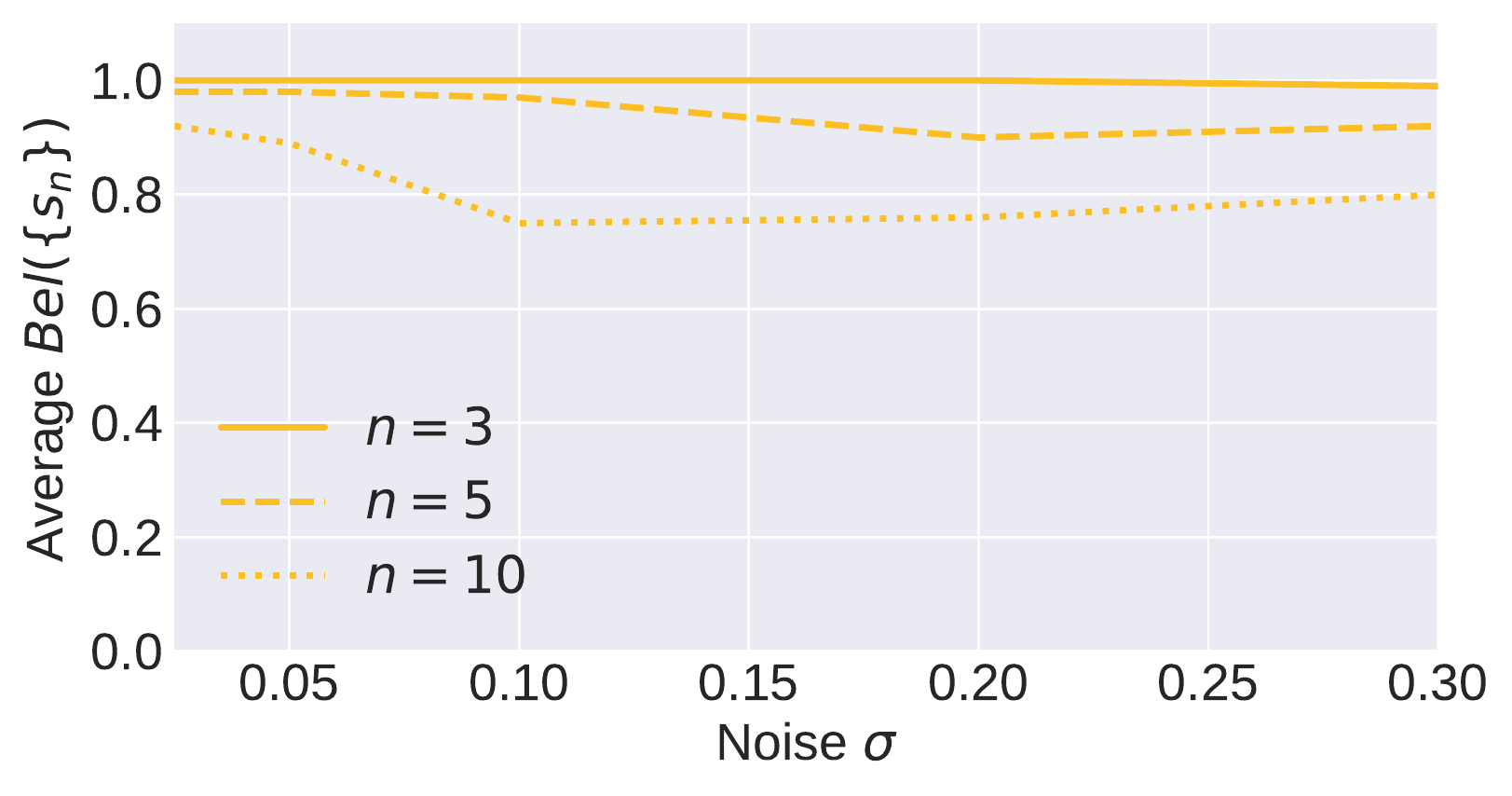}
    \subcaption{Yager's rule.}
    \label{fig:best-of-n-yagers}
\end{subfigure}
\caption{Average $Bel(\{s_n\})$ for $n \in \{3,5,10\}$ plotted against $\sigma$ for $r = 0.05$.}
\label{fig:best-of-n}
\end{figure}

Noise is ubiquitous in applications of multi-agent systems. In embodied agents such as robots this is often a result of sensor errors, but noise can also be a feature of an inherently variable environment. In this section we consider the effect of evidential noise on the best-of-$n$ problem, as governed by the standard distribution $\sigma$ of the noise.

Figure~\ref{fig:noise} shows the average value of $Bel(\{s_3\})$ at steady state plotted against $\sigma \in [0, 0.3]$ for different evidence rates $r \in \{0.01, 0.05, 0.1\}$. Figure~\ref{fig:noise} (left) shows that for an evidence rate $r = 0.01$, all operators except the averaging operator have very similar performance in the presence of noise.
For example with no noise, i.e.,\ $\sigma = 0$, Yager's rule converges to an average value of $0.97$, Dubois \& Prade's operator converges to an average of $Bel(\{s_3\}) = 0.96$, Dempster's rule to $0.95$ on average, and the averaging operator to $0.4$. Then, with $\sigma = 0.3$, Yager's rule converges to an average value of $0.8$, Dubois \& Prade's operator to an average value of $Bel(\{s_3\}) = 0.77$, Dempster's rule to $0.74$, and the averaging operator converges to $0.29$. Hence, all operators are affected by the noise to a similar extent given this low evidence rate.

In contrast, for the evidence rates of $r=0.05$ and $r=0.1$, Figures~\ref{fig:noise} (centre) and \ref{fig:noise} (right), respectively, we see that both Dubois \& Prade's operator and Yager's rule are the most robust combination operators to increased noise.
Specifically, for $r=0.05$ and $\sigma = 0$, they both converge to an average value of $Bel(\{s_3\}) = 1$ and for $\sigma = 0.3$ they only decrease to $0.99$. On the other hand, the presence of noise at this evidence rate has a much higher impact on the performance of Dempster's rule and the averaging operator. For $\sigma = 0$ Dempster's rule converges to an average value of $Bel(\{s_3\}) = 0.95$ but this decreases to $0.78$ for $\sigma = 0.3$, and for the averaging operator the average value of $Bel(\{s_3\}) = 0.41$ and decreases to $0.29$. The contrast between the performance of the operators in the presence of noise is even greater for the evidence rate $r=0.1$ as seen in Figure~\ref{fig:noise} (right). However, both Dubois \& Prade's operator and Yager's rule differ in this context since, for both evidence rates $r=0.05$ and $r=0.1$, their average values of $Bel(\{s_3\})$ remain constant at approximately $1$.


\subsection{Scalability to Larger Numbers of States}
\label{ssec:scalability}


In the swarm robotics literature most best-of-$n$ studies are for $n=2$ (see for example~\cite{Valentini2014,Reina2018}). However, there is a growing interest in studying larger numbers of choices in this context~\cite{Reina2017,Crosscombe2017}.
Indeed, for many distributed decision-making applications the size of the state space, i.e., the value of $n$ in the best-of-$n$ problem, will be much larger.
Hence, it is important to investigate the scalability of the proposed DST approach to larger values of $n$.

Having up to now focused on the $n=3$ case, in this section we present additional simulation results for $n=5$ and $n=10$. As proposed in Section~\ref{subsection:parameters}, the quality values are allocated so that $q_i=\frac{i}{n+1}$ for $i=1, \ldots, n$.
Here, we only consider Dubois \& Prade's operator and Yager's rule due to their better performance when compared with the other two combination operators.

Figure~\ref{fig:best-of-n} shows the average values of $Bel(\{s_n\})$ at steady state plotted against noise $\sigma \in [0,0.3]$ for evidence rate $r=0.05$, where $Bel(\{s_n\})$ is the belief in the best state for $n=3,\ 5$ and $10$.
For Dubois \& Prade's operator, Figure~\ref{fig:best-of-n-consensus} shows the steady state values of $Bel(\{s_3\})=1$ independent of the noise level, followed closely by the values of $Bel(\{s_5\})=0.94$ at $\sigma=0$ for the $n=5$ case. However, for $n=10$ the value of $Bel(\{s_{10\}})$ is $0.61$ when $\sigma=0$, corresponding to a significant decrease in performance. At the same time, from Figure~\ref{fig:best-of-n-yagers}, we can see that for Yager's rule performance declines much less rapidly with increasing $n$ than for Dubois \& Prade's operator.
So at $\sigma=0$ and $n=5$ the average value at steady state for Yager's rule is almost the same as for $n=3$, i.e. $Bel(\{s_5\}) = 0.98$, with a slight decrease in the performance $Bel(\{s_{10}\})=0.92$ for $n=10$.
As expected the performance of both operators decreases as $\sigma$ increases, with Yager's rule being much more robust to noise than Dubois \& Prade's operator for large values of $n$.



In this way, the results support only limited scalability for the DST approach to the best-of-$n$ problem, at least as far as uniquely identifying the best state is concerned. Furthermore, as $n$ increases so does sensitivity to noise. This reduced performance may in part be a feature of the way quality values have been allocated. Notice that as $n$ increases, the difference between successive quality values $q_{i+1}-q_i=\frac{1}{n+1}$ decreases.
This is likely to make it difficult for a population of agents to distinguish between the best state and those which have increasingly similar quality values.
Furthermore, a given noise standard deviation $\sigma$ results in an inaccurate ordering of the quality values the closer those values are to each other, making it difficult for a population of agents to distinguish between the best state and those which have increasingly similar quality values.




\section{Conclusions and Future Work}
\label{section:conclusion}

In this paper we have introduced a model of consensus formation in the best-of-$n$ problem which combines updating from direct evidence with belief combination between pairs of agents.
We have utilised DST as a convenient framework for representing agents' beliefs, as well as the evidence that agents receive from the environment.
In particular, we have studied and compared the macro-level convergence properties of several established operators applied iteratively in a dynamic multi-agent setting and through simulation we have identified several important properties of these operators within this context.
Yager's rule and Dubois \& Prade's operator are shown to be most effective at reducing polarisation and reaching a consensus for all except very low evidence rates, despite them not satisfying certain desirable properties, e.g., Dubois \& Prade's operator is not associative while Yager's rule is only quasi-associative~\cite{Dubois2016}. Both have also demonstrated robustness to different noise levels. However, Yager's rule is more robust to noise than Dubois \& Prade's operator for large values of states $n>3$. Although the performance of both operators decreases with an increase in the number of states $n$, Yager's rule is shown to be more scalable.
We believe that underlying the difference in the performance of all but the averaging operator is the way in which they differ in their handling of inconsistent beliefs. Specifically, the manner in which they reallocate the mass associated with the inconsistent non-overlapping sets in the case of Dempster's rule, Dubois \& Prade's operator and Yager's rule.


%


Further work will investigate the issue of scalability in more detail, including whether alternatives to the updating process may be applicable in a DST model, such as that of negative updating in swarm robotics~\cite{Lee2018B}. We must also consider the increasing computational cost of DST as the size of the state space increases and investigate other representations such as possibility theory~\cite{Dubois2001} as a means of avoiding exponential increases in the cost of storing and combining mass functions. Finally, we hope to adapt our method to be applied to a network, as opposed to a complete graph, so as to study the effects of limited or constrained communications on convergence.

\section*{Acknowledgments}
This work was funded and delivered in partnership between Thales Group, University of Bristol and with the support of the UK Engineering and Physical Sciences Research Council, ref.\ EP/R004757/1 entitled ``Thales-Bristol Partnership in Hybrid Autonomous Systems Engineering (T-B PHASE).''

Thanks to Palina for helping to identify a fundamental issue with an earlier version of this work which has since been corrected in the present version.

\bibliographystyle{named}
\bibliography{references}

\end{document}